\documentclass[english,aip,preprint,jcp]{revtex4-1}

\makeatletter   
\usepackage{bm}
\usepackage[colorlinks=true,linkcolor=blue]{hyperref}

\usepackage{amsmath,amssymb}
\usepackage{graphicx}
\usepackage{caption}
\usepackage{subcaption}
\usepackage{color}
\usepackage{dcolumn}
\usepackage{babel}
\usepackage{natbib}

\begin{document}
\title{Charge transfer excitations from excited state Hartree-Fock
subsequent minimization scheme}

\author{Iris Theophilou}
\email[Corresponding author:]{i.theophilou@fz-juelich.de }
\affiliation{Peter Grunberg Institut (PGI) Forschungszentrum J\"ulich, D-52425, J\"ulich, Germany}
\author{M. Tassi}
\affiliation{Institute for Advanced Materials, Physicochemical Processes, Nanotechnology and Microsystems, 'Demokritos' National Center for
Scientific Research, 15310, Athens, Greece}
\author{S. Thanos}
\affiliation{Institute for Advanced Materials, Physicochemical Processes, Nanotechnology and Microsystems, 'Demokritos' National Center for
Scientific Research, 15310, Athens, Greece}

\begin{abstract}

Photoinduced charge-transfer processes play a key role for novel
photovoltaic phenomena and devices. Thus, the development of ab initio
methods that allow for an accurate and computationally inexpensive treatment of
charge-transfer excitations is a topic that nowadays attracts a lot of
scientific attention. In this paper we extend an approach recently
introduced for the description of single and double excitations (M. Tassi,
I. Theophilou and S. Thanos, Int. J. Quantum Chem., \textbf{113}, 690
(2013), M. Tassi, I. Theophilou and S. Thanos, J. Chem. Phys. \textbf{138},
124107 (2013)) to allow for the description of intermolecular charge-transfer excitations.
We describe an excitation where an electron is transferred from a donor system to an acceptor one,
keeping the excited state orthogonal to the ground state and avoiding
variational collapse. These conditions are achieved by decomposing the
space spanned by the Hartree-Fock (HF) ground state orbitals into four
subspaces: The subspace spanned by the occupied orbitals that are localized
in the region of the donor molecule, the corresponding for the acceptor ones
and two more subspaces containing the virtual orbitals that are localized in
the neighborhood of the donor and the acceptor, respectively. Next, we create
a Slater determinant with a hole in the subspace of occupied orbitals of the
donor and a particle in the virtual subspace of the acceptor. Subsequently
we optimize both the hole and the particle by minimizing the HF energy
functional in the corresponding subspaces. Finally, we test our approach by
calculating the lowest charge-transfer excitation energies for a set of
tetracyanoethylene-hydrocarbon complexes, that have been used earlier as a
test set for such kind of excitations.
\end{abstract}
\maketitle

\section{INTRODUCTION}

Recently, organic photovoltaics have attracted a great deal of attention, as they
have the potential of harvesting solar light cheaply and easily\cite%
{ChemRev1}. For this purpose a lot of scientific effort has been devoted to
simulating some of the fundamental steps occurring in natural
photosynthesis, one of the most important being the photoinduced charge
separation\cite{ChemRev2}. Thus, the scientific community adopted as one of its major tasks 
to develop methods that can reliably and inexpensively describe
charge-transfer (CT) excitations, where light absorption causes a charge
transfer from a donor to an acceptor. The donor-acceptor system consists of 
 either two or more different molecules that interact
weakly or of two sites of the same big molecule. In this work we shall deal
only with the first class of donor-acceptor complexes.

Although time-dependent density-functional theory (TDDFT)\cite{TDDFT1, TDDFT2, TDDFT3} is currently successfully
applied for the description of excitations of large molecular systems, charge-transfer excitations in the linear response regime with
standard frequency-independent exchange-correlation kernels exhibit significant
failures, giving underestimations of several eV \cite{Head-Gordon, Tozer}.
This failure is attributed to the wrong asymtotic shape of the ground-state exchange correlation (xc) potential,
which leads to errors in the orbital energies\cite{TDDFTbookCT}.
Moreover, the Kohn-Sham orbital overlaps of the donor$-$%
acceptor system that enter the exchange-correlation part of the Dyson
equation vanish when non-hybrid xc kernels are used. Consequently, the charge-transfer energy is given as the orbital energy difference\cite{Peach}. Currently,
there is a lot of effort on the development of exchange correlation kernels
that mitigate this deficiency\cite{Tawada,Gritsenko, Lange, Hesselmann,
Stein}. Using range-separated hybrid functionals \cite%
{Tawada,Stein}, where the exchange functional is split in two terms, a
short-range term that is represented by a local density approximation (LDA) or generalized gradient approximation (GGA) potential and a
long-range term that is treated via exact exchange, improve the
results obtained by TDDFT for charge-transfer
excitations. In this case one has to determine the range-splitting
parameter. This is done either empirically by fitting to a set of data \cite%
{Lange} or by introducing a system dependent parameter by tuning the highest occupied molecular orbital (HOMO)
 close to minus the ionization potential\cite{Stein}. A
time-independent variational approach of density functional theory (DFT), that is not based on response
theory, termed as relaxed constrained-variational DFT, gives promising results
on charge-transfer excitations\cite{Ziegler}. In this case, the charge
constraint is imposed on a region of the orbital space. There is also a
constrained-variational DFT scheme where the charge is constrained in real
space \cite{Van Voorhis}. Recently, a perturbative-delta self consistent field ($\Delta $SCF)\cite{perdscf}
approach has been developed, where a non-aufbau occupation of the density
during a normal self consistent field optimization is enforced as in
traditional $\Delta $SCF\cite{dscf}, which in combination with perturbation
theory, gives for the systems tested a good description of excitations in
donor-acceptor complexes. \par
\begin{figure}
\includegraphics[width=0.4\textwidth]{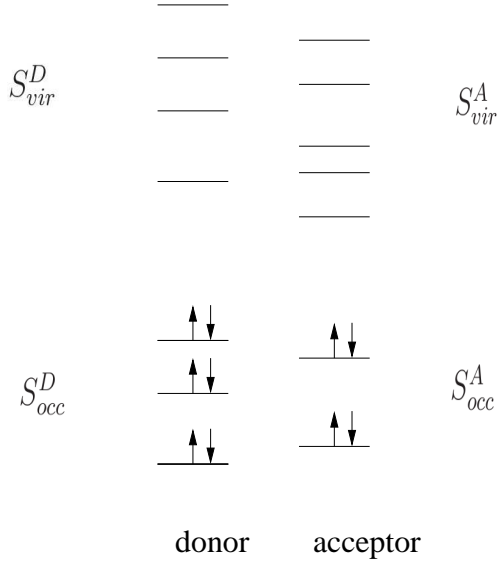}
\caption{Sketch of the subspaces $S_{occ}^{D}$ (occupied orbitals of the donor),  $S_{occ}^{A}$ (occupied orbitals of the acceptor),  $S_{vir}^{D}$ (virtual orbitals
of the donor) and $S_{vir}^{A}$ (virtual orbitals of the acceptor). Note that in practice we perform the subspace decomposition only for the up orbitals and the down electrons are just given here to complete the picture.}
\label{figure_subspaces}
\end{figure}

Recently, a variational approach was introduced, based on solving the
Unrestricted Hartree-Fock (UHF)\cite{UHF} eigenvalue equations in different
subspaces spanned by the occupied and virtual ground state orbitals\cite%
{our1}. In this approach, an excited state was considered with a hole in the
subspace of the occupied orbitals and a particle in the subspace of virtual
orbitals, where both were determined variationally. Conceptually, this
scheme is close to the $\Delta $SCF one, however within this scheme there is
no possibility of a variational collapse to the ground state, as it may
happen in $\Delta $SCF, since the state defined in this way, is always
orthogonal to the ground state. The orthogonality of approximate excited states to the ground state is desirable since the exact eigenstates of the Hamiltonian also have this property.
  In a recent paper an extension of this approach was given which includes double excitations\cite{our2}. In the
current work, in order to study an excitation where an electron is
transferred from a donor system to an acceptor one, it is necessary to take
into account not only orthogonality, but also the localization of the orbitals
in the donor and acceptor regions, with some overlap in the intermediate
region. For this purpose we create four subspaces: one that belongs to the
occupied orbitals that are localized mostly in the donor region and a
similar one consisting of the virtual ones. A third set consists of the occupied
orbitals localized in the neighborhood of the acceptor and a fourth set is
spanned by the virtual orbitals of the acceptor. See Fig.(\ref{figure_subspaces}) for an example where the donor side contains
three occupied orbitals and four unoccupied ones in the respective subspaces, and the acceptor side provides two occupied
and five unoccupied orbitals. 
Next, we describe the CT excited
state by a Slater determinant with a hole in the donor occupied subspace and
a particle in the acceptor virtual subspace. Both, the particle and the
hole, are determined variationally by minimizing the Hartree Fock
functional. In this way we are able to calculate intermolecular charge-transfer excitations. More details are given in the next section. We
applied our scheme to a series of Tetracyanoethylene (TCNE)-various donors systems that have been
used before \cite{Stein, Ziegler,perdscf, gw} and found that our results
compare well with the experimental ones. This paper is organized
as follows: In section II we explain the main idea of our approach and give the equations used for its realization. In section III we examine the
performance of this scheme in obtaining charge-transfer excitation energies
and we show that at least for the system tested, it can provide a good
description. Finally, in section IV we give our
concluding remarks.

\section{Restrictions to the charge-transfer excited state orbitals}

Let us consider an excitation of one electron from the donor to
the acceptor. We assume that the excited electron is a spin up one
and for simplifying our notation we shall omit the spin index when we refer
to spin up. All the orbitals derived from the Hartree-Fock ground state
calculation, which will be denoted by $\varphi_{j}$, are used to build two
orthogonal subspaces, $S^{D},$ consisting of the orbitals $\varphi_{j}^{D}$
 that are localized mostly in the donor and $S^{A}$ of orbitals similarly 
localized mostly in the acceptor system $\varphi_{j}^{A}$. This classification
 is performed  by summing for each orbital $\varphi _{j}$, its projections on the gaussians
centered at the donor or correspondingly at the acceptor. Thus, in order to
check whether the orbital $\varphi _{i}$ is more localized in the area of
the donor or the acceptor, we calculate the sum of the overlaps $\sum_{k}{%
|\langle g_{k}|\varphi _{i}\rangle |}^{2}$ with $k$ runing over all
gaussians $|g_{k}\rangle $ centered in the donor or the acceptor
system. Then, depending on which sum is larger, we attribute the orbital $%
\varphi _{i}$ to the donor or the acceptor subspace. When donor and
acceptor are not molecules of the same type an orbital cannot have exactly
equal projections on both the donor and the acceptor. Next, we further
separate $S^{D}$ into a subspace $S_{occ}^{D}$ spanned by the occupied orbitals \{$%
\varphi _{1}^{D},..\varphi _{N_{occ}^{D}}^{D}\}$ and $S_{vir}^{D}$ of \ the
virtual ones $\{\varphi _{{N_{occ}+1}^{D}}^{D},..,\varphi _{N^{D}}^{D}\}$.
Similarly $S^{A}$ consists of $S_{occ}^{A}$ spanned $\{\varphi
_{1}^{A},..\varphi _{N_{occ}^{A}}^{A}\}$ and $S_{vir}^{A}$ of the virtuals of
the acceptor $\{\varphi _{{N_{occ}+1}^{A}}^{A},..,\varphi _{N^{A}}^{A}\}$. The
subspace dimensions are denoted by $N_{{}}^{D}$ for $S^{D},$ $N_{occ}^{D}$
for $S_{occ}^{D}$ and correspondingly by $N^{A},$ $N_{occ}^{A}$ for the acceptor. 
The charge-transfer excited state orbitals that are localized in the donor system are
denoted by $\chi _{i}^{D}$ and the ones that are localized in the acceptor
by $\chi _{i}^{A}$. Then, we demand that the excited state $|\Phi
_{CT}\rangle =|\chi _{1}^{D},...\chi _{N_{occ}^{D}-1}^{D},\chi
_{1}^{A}...\chi _{{N_{occ}^{A}}+1}^{A},\chi _{1}^{\downarrow },...\chi
_{N\downarrow }^{\downarrow }\rangle $ has one more electron in $S_{vir}^{A}$
than the ground state HF and one hole in $S_{occ}^{D}$. Thus, in addition to the
normalization of the orbitals the following conditions should be satisfied: 
\newline
i) $\langle \chi _{i}^{D}|\varphi _{j}^{A}\rangle =0$ for $i\leq
N_{occ}^{D}-1$ and $j=1,..,N^{A}$\newline
ii) $\langle \chi _{i}^{D}|\varphi _{j}^{D}\rangle =0$ , for $i\leq
N_{occ}^{D}-1$ and $j\geq N_{occ}^{D}+1$\newline
The two relations above imply that the excited orbitals $\chi
_{1}^{D},...\chi _{N_{occ}^{D}-1}^{D}$ belong to the subspace $S_{occ}^{D}$.%
\newline
iii) $\langle \chi _{{N_{occ}^{A}}+1}^{A}|\varphi _{j}^{A}\rangle =0$ , for $%
j\leq {N_{occ}^{A}}$ \newline
iv)$\langle \chi _{{N_{occ}^{A}}+1}^{A}|\varphi _{j}^{D}\rangle =0$ , for $%
j=1,..,N^{D}$\newline
Relations (iii) and (iv) imply that $|\Phi _{CT}\rangle $ comprises one orbital named $\chi _{{N_{occ}^{A}}+1}^{A}$ that belongs
to $S_{vir}^{A}$.\newline
v) $\langle \chi _{i}^{A}|\varphi _{j}^{A}\rangle =0$ , for $i\leq
N_{occ}^{A}$ and for $j>{N_{occ}^{A}}$ and \newline
vi)$\langle \chi _{i}^{A}|\varphi _{j}^{D}\rangle =0$ , for $i\leq
N_{occ}^{A}$ and for j=1,..,$N^{D}$\newline
Relations (v) and (vi) imply that the excited orbitals $\chi
_{1}^{A},..,\chi _{N_{occ}^{A}}^{A}$ belong to $S_{occ}^{A}$. \newline
Minimizing the energy functional $\langle \Phi _{CT}|\hat{H}|\Phi
_{CT}\rangle $ under the above conditions, taking also into account the orbital normalization, one gets
the following set of one particle equations:

\vspace{1pt} 
\begin{eqnarray}
\hat{F}|\chi _{i}^{D}\rangle -\sum_{j\leq N^{A}}|\varphi _{j}^{A}\rangle
\langle \varphi _{j}^{A}|\hat{F}|\chi _{i}^{D}\rangle
-\sum_{j>N_{occ}^{D}}|\varphi _{j}^{D}\rangle \langle \varphi _{j}^{D}|\hat{F%
}|\chi _{i}^{D}\rangle =\epsilon _{i}^{D}|\chi _{i}^{D}\rangle
,\quad \text{for $i<N_{occ}^{D}$}
\label{donor}
\end{eqnarray}%
 As usual, the Fock
operator $\hat{F}$ corresponding to $\Phi _{CT}$ is defined by the following
equation 
\begin{equation}
(\hat{F}^{s}\chi _{i}^{s})(\mathbf{r})=\hat{h}\chi _{i}^{s}(\mathbf{r})+\int
d\mathbf{r^{\prime }}\frac{\rho (\mathbf{r^{\prime }};\Phi _{CT})}{|\mathbf{r%
}-\mathbf{r^{\prime }}|}\chi _{i}^{s}(\mathbf{r})-\int d\mathbf{r^{\prime }}%
\frac{\rho _{s}(\mathbf{r,r^{\prime }};\Phi _{CT})}{|\mathbf{r}-\mathbf{%
r^{\prime }}|}\chi _{i}^{s}(\mathbf{r^{\prime }}),
\end{equation}%
where $\hat{h}(\mathbf{r})=-\frac{1}{2}\nabla ^{2}+\hat{V}(\mathbf{r})$ is
the kinetic plus external potential operator, which is the same as the one
of the UHF ground state and $\rho_{s} (\mathbf{r},\mathbf{r^{\prime }};\Phi _{CT})$ is the spin density matrix of $|\Phi _{CT}\rangle $: 
\begin{equation}
\rho _{s}(\mathbf{r,r^{\prime }};\Phi _{CT})=\sum_{i=1}^{N^{s}}\chi _{i}^{s}(%
\mathbf{r)}\chi _{i}^{s}(\mathbf{r^{\prime }}),
\end{equation}%
where $N_{s}$ stands for the number of occupied orbitals with $s$ $\uparrow $ or $%
\downarrow $. Finally, $\rho (\mathbf{r};\Phi _{CT})=\rho _{\uparrow }(%
\mathbf{r,r};\Phi _{CT})+\rho_{\downarrow}(\mathbf{r,r};\Phi_{CT})$. We use the notation $(\hat{F}^{s}\chi _{i}^{s})(%
\mathbf{r})$ to show that $\hat{F}^{s}$ maps a function and not its value to
another function in order to account for the nonlocal operator.
The sums $\sum_{j>N_{occ}^{D}}|\varphi _{j}^{D}\rangle \langle \varphi _{j}^{D}|$
and $\sum_{j\leq N^{A}}|\varphi _{j}^{A}\rangle \langle \varphi
_{j}^{A}| $ are projection operators onto the subspaces $%
S_{vir}^{D}$ and $S^{A}$ respectively, which act as identity operators in
the corresponding subspaces. For the excited orbitals of the acceptor we obtain:\newline
\vspace{1pt} 
\begin{eqnarray}
\hat{F}|\chi _{i}^{A}\rangle -\sum_{j\leq N^{D}}|\varphi _{j}^{D}\rangle
\langle \varphi _{j}^{D}|\hat{F}|\chi _{i}^{A}\rangle -\sum_{j\leq
N_{occ}^{A}}|\varphi _{j}^{A}\rangle \langle \varphi _{j}^{A}|\hat{F}|\chi
_{i}^{A}\rangle =\epsilon _{i}^{A}|\chi _{i}^{A}\rangle ,\quad \text{for $i>N_{occ}^{A}$}
\label{acceptor1}
\end{eqnarray}%
\vspace{1pt} and 
\begin{eqnarray}
\hat{F}|\chi _{i}^{A}\rangle -\sum_{j\leq N^{D}}|\varphi _{j}^{D}\rangle
\langle \varphi _{j}^{D}|\hat{F}|\chi _{i}^{A}\rangle -\sum_{j>N_{occ}^{A}}|\varphi
_{j}^{A}\rangle \langle \varphi _{j}^{A}|\hat{F}|\chi _{i}^{A}\rangle
=\epsilon _{i}^{A}|\chi _{i}^{A}\rangle ,\quad \text{for $i\leq N_{occ}^{A}$}
\label{acceptor2}
\end{eqnarray}%
The first sum that appears in Eqs.(\ref{acceptor1}), (\ref{acceptor2})
corresponds to the projection operator in $S^{D}$ and the second
sum to projections in $S_{occ}^{A}$ and $S_{vir}^{A}$, respectively. Then
the left hand side of Eq.(\ref{donor}) can be identified as the projection of
the Fock operator onto the subspace of the occupied orbitals of the donor, Eq.%
(\ref{acceptor1}) as the projection of the Fock operator in the virtual orbitals
of the acceptor and Eq.(\ref{acceptor2}) as the projection of the Fock operator in the
occupied orbitals of the acceptor. Note that we need no equation for obtaining orbitals
that belong to $S^{D}_{vir}$ since these orbitals do not enter our $|\Phi_{CT}\rangle$ state. \par
The equation for the spin down orbitals has the usual UHF form\cite{UHF} i.e.
\begin{eqnarray}
\hat{F}^{\downarrow }|\chi _{i}^{\downarrow }\rangle =\epsilon
_{i}^{\downarrow }|\chi _{i}^{\downarrow }\rangle  \label{down}
\end{eqnarray}
since no constraints are imposed beyond that of normalization. Note, however,
that these orbitals are not identical to those of the UHF ground state since
the Hartree term changes because it includes the density of the spin up
particles. Since we assumed that the electron which gives the charge-transfer 
excited state is an up electron, the Fock matrices in Eqs.(\ref%
{donor}),(\ref{acceptor1}),(\ref{acceptor2}) are Fock matrices of spin up electrons. 

As mentioned earlier, the left hand side of Eqs.(\ref{donor}), (\ref%
{acceptor1}), (\ref{acceptor2}) is nothing more than the projection of the Fock
matrices in the subspaces $S_{occ}^{D}$, $S_{vir}^{A}$ and $S_{occ}^{A}$
respectively, forcing our one electron eigenvalue equations to have
solutions that belong to these subspaces, making it possible to avoid the use of
Lagrange multipliers. This is achieved by constructing the
Fock matrices in the basis of the ground state HF orbitals that span the
corresponding subspaces. For more details see Appendix B of \cite%
{our2}. Once we manage to restrict the Fock operator in the subspaces $%
S_{occ}^{D}$, $S_{vir}^{A}$ and $S_{occ}^{A}$,  Eqs.(\ref{donor}),(\ref%
{acceptor1}),(\ref{acceptor2}) assume the following simple form: 
\begin{eqnarray}
\hat{F}_{occ}^{D}|\chi _{i}^{D}\rangle &=&\epsilon _{i}^{D}|\chi
_{i}^{D}\rangle  \label{eq:eigenstate donor}, \\
\hat{F}_{vir}^{A}|\chi _{i}^{A}\rangle &=&\epsilon _{i}^{A}|\chi
_{i}^{A}\rangle  \label{eq:eigenstate acceptor1}, \\
\hat{F}_{occ}^{A}|\chi _{i}^{A}\rangle &=&\epsilon _{i}^{A}|\chi
_{i}^{A}\rangle  \label{eq:eigenstate acceptor2}.
\end{eqnarray}%
From the $N^{D}-1$ lowest energy solutions of Eq.(\ref{eq:eigenstate
donor}), the lowest energy solutions of Eq.(\ref{eq:eigenstate acceptor1}) and
all the solutions of Eq.(\ref{eq:eigenstate acceptor2}) we can find the $%
N^{D}-1$ $\chi _{i}^{D}$, the $\chi _{N_{occ}^{A}+1}^{A}$ and the $\chi
_{1}^{A},...,\chi _{N_{occ}^{A}}^{A}$ orbitals of $|\Phi _{CT}\rangle $.
Thus, one has to self consistently solve the system of Eqs.(\ref%
{eq:eigenstate donor}),(\ref{eq:eigenstate acceptor1}),(\ref{eq:eigenstate
acceptor2}) together with the equation for the down orbitals, Eq.(\ref{down}). The
determinant $|\Phi _{CT}\rangle $ that corresponds to the approximate
eigenstate of the excited charge-transfer state will be orthogonal to the
ground state one, i.e. $\langle \Phi _{0}|\Phi _{CT}\rangle =0$. This is
true since the highest occupied orbital of the acceptor in $|\Phi
_{CT}\rangle $ belongs to $S_{vir}^{A}$, thus this orbital by construction
has zero overlap with all the ground state occupied orbitals.

The next step is to express the Fock operators that appear in the eigenvalue
equations as matrices in some basis set to do the appropriate numerical
calculations. Since any basis set we use to do this expansion will lead to
an eigenvalue problem giving the same results (as long as the basis is
complete), we can use the ground state orbitals that span the subspaces $%
S^{D}_{occ}$, $S^{A}_{vir}$ and $S^{A}_{occ}$ to express the Fock matrices
of equations (\ref{eq:eigenstate donor}), (\ref{eq:eigenstate acceptor1}), (\ref%
{eq:eigenstate acceptor2}). 
\begin{equation}
F_{occ,i,j}^{ D}=\langle \varphi _{i}|\hat{F}|\varphi _{j}\rangle 
\text{ where }\varphi _{i},\varphi _{j}\:\epsilon\: S_{occ}^{D},
\end{equation}%
\begin{equation}
F_{vir,i,j}^{A}=\langle \varphi _{i}|\hat{F}|\varphi _{j}\rangle 
\text{ where }\varphi _{i},\varphi _{j}\:\epsilon\: S_{vir}^{A},
\end{equation}%
\begin{equation}
F_{occ,i,j}^{ A}=\langle \varphi _{i}|\hat{F}|\varphi _{j}\rangle 
\text{ where }\varphi _{i},\varphi _{j}\:\epsilon \: S_{occ}^{A}.
\end{equation}%
\qquad The equations above are the ones that we solve in order to obtain $|\Phi_{CT}\rangle$.
It is worth to note that one could also treat charge-transfer
excitations where more than one electrons are transferred from the donor to
the acceptor system. The only difference in the treatment is to introduce
two or more excited state orbitals in $S_{vir}^{A}$ and to create two or
more holes respectively in the subspace of $S_{occ}^{D}$.

We stress the fact that all orbitals that contribute to $\Phi _{CT}$ are
obtained by one electron eigenvalue equations where they are "repelled"
correctly by an electrostatic charge of $N-1$ electrons. This is in contrast
to HF virtual orbitals which are artificially diffuse as they are repelled
by $N$ electrons. Consequently, occupying the HF virtual orbitals without further
minimization, as it is well known, leads to large excitation energies.

\section{RESULTS AND DISCUSSION}

The approach developed in section II was used to
calculate the charge-transfer excitation energies of a test set of various
aromatic donor - TCNE complexes and Anthracene substituted derivatives -
TCNE, for which experimental results are available. This test set was
introduced by Stein et al. \cite{Stein} to examine the
performance of their dual-range functional in charge-transfer excitations.
Thereafter, it has been used as a standard set by other authors to validate
their approaches in this kind of excitations \cite{Ziegler, perdscf, gw}.
For all the donor-TCNE results obtained by using the approach presented in
this work, the 6-31G* basis set was adopted. The geometries used
were the B3LYP-optimized ones and were taken from Stein et al\cite{Stein}.

\begin{table}[tbp]
\caption{Charge transfer excitation energies (eV) for $\protect\pi$ donor to 
$\protect\pi^{*}$ (TCNE) transitions in donor-TCNE complexes in gas phase}
\label{tbl1}%
\begin{minipage}{\textwidth}
\begin{tabular}{c c c c c c }\hline \hline
Donor & TDDFT-B3LYP \footnote{Taken from Ref.(\cite{Stein})}& TDDFT-BNL$\gamma^{*}$\footnote{Range-split BNL functional from
 Ref.(\cite{Stein}) } 
& $\Delta$SCF-HF & This work & Exp\footnote{Experimental gas-phase data from Ref. (\cite{exp1})} \\ \hline
Benzene & 2.1 & 3.80   & 3.06 & 3.46 & 3.59 \tabularnewline
Toluene & 1.8 & 3.40   & 2.75 & 3.16 & 3.36 \tabularnewline
O-xylene &1.5  & 3.00    & 2.56 & 2.89 & 3.15 \tabularnewline
Naphtalene & 0.9  & 2.70   & 2.21 & 2.61 & 2.60 \tabularnewline
\hline \hline
\end{tabular}
\end{minipage}
\end{table}

\clearpage
\begin{table}[tbp]
\caption{Charge transfer excitation energies (eV) for $\protect\pi $ donor
to $\protect\pi ^{\ast }$ transitions in Substituted Anthracene - TCNE
complexes in solution}
\label{tbl2}%
\begin{minipage}{\textwidth}
\begin{tabular}{c c c c c c c c}\hline \hline
Substituent &TDDFT-PBE \footnote{Taken from Ref. (\cite{Stein})}& TDDFT-B3LYP \footnote{Taken from Ref.(\cite{Stein})}& TDDFT-BNL$\gamma^{*}$\footnote{Range-split BNL functional from
 Ref.(\cite{Stein}) }  
& $\Delta$SCF-HF & This work & Exp\footnote{Experimental solution data from Ref. (\cite{exp2}) with CH3Cl as solvent} \\ \hline
None & 0.9 & 1.00  & 1.82& 1.63 & 1.60 & 1.73 \tabularnewline
9-cyano & fail & 0.5  & 2.03&1.36   & 2.00 & 2.01 \tabularnewline
9-chloro &0.9  & 1.0  & 1.82&1.57   & 1.72 & 1.74 \tabularnewline
9-carbo-methoxy & 0.8  & 0.9   & 1.74 & 1.38 & 1.80 & 1.84 \tabularnewline
9-methyl &1.0  & 1.1  & 1.71 &1.37 & 1.40 & 1.55 \tabularnewline
9,10-dimethyl &1.3  & 1.4  & 1.77 & 1.57 & 1.36 & 1.44 \tabularnewline
9-formyl & 0.8  & 1.0  & 1.95   &1.64 & 1.90 & 1.90 \tabularnewline
9-formyl 10-chloro &0.8  & 0.9  & 1.96 &1.70  &  2.03 & 1.96 \tabularnewline
\hline \hline
\end{tabular}
\end{minipage}
\end{table}
In Table \ref{tbl1}, we present the gas phase excitation energies from
various aromatic donors to TCNE. For comparison, we also give the
experimental values of these excitations and the results obtained by
applying TDDFT-B3LYP, TDDFT dual-range BNL $\gamma ^{\ast }$ taken from
 reference \cite{Stein} and $\Delta $SCF-HF. $\Delta $SCF-HF results are obtained by starting with the UHF orbitals of the ground-state as initial guess and then solving the UHF equations using a SCF procedure in which the lowest $N^{\uparrow}$-1 orbitals and the $(N^{\uparrow}+1)^{th}$ are occupied in each update of the density.
 As one can see, all the TDDFT-B3LYP results for these
systems underestimate the experimentally measured excitation
energies by about 1.5 eV, $\Delta $SCF-HF results are better but still underestimate the energy by 0.4-0.6 eV, whereas those of the present approach
are in good agreement with experiment and those obtained by TDDFT-BNL $%
\gamma ^{\ast }$. 

In Table \ref{tbl2} we give the results obtained for Anthracene substituted
derivatives-TCNE in methylene chloride solution. As no solvation model was
used, 0.32 eV were substracted from every calculated gas phase value,
following the suggestion by Stein et al\cite{Stein}. Thus all the calculated results
that appear in Table \ref{tbl2} are the calculated values minus 0.32 eV to
account for the solvent effect. For comparison we give the TDDFT-PBE,
TDDFT-B3LYP, TDDFT dual-range BNL $\gamma ^{\ast }$ and the experimental
results for these excitations. The TDDFT-PBE and TDDFT-B3LYP results are not
only too low, but also the descending order relation of the various donor-acceptor
excitations is violated. The descending order relation is also violated by $\Delta$SCF-HF results. Our results keep the descending order, are close to TDDFT-BNL$\gamma ^{\ast }$
and compare well to experiment.\par
As expected from the self-interction problem of virtual HF orbitals, replacing in the ground state Slater determinant the HOMO of the donor with the lowest unoccupied molecular orbital (LUMO) of the acceptor and calculating the energy difference from the ground state HF energies gave more than 1eV overestimation of the CT energies for all the systems tested.\par
\begin{figure}
\includegraphics[width=0.45\textwidth]{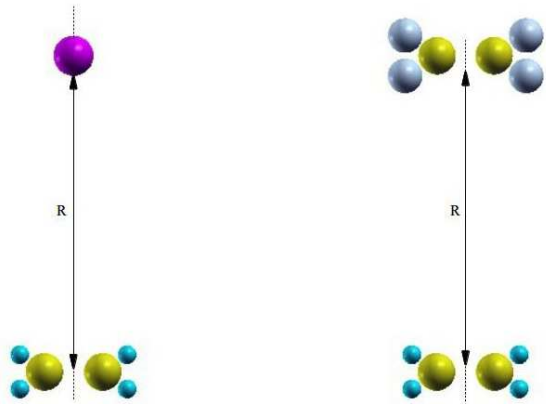}
\caption{Structures of C$_{2}$H$_{4}$-Ne and C$_{2}$H$_{4}$-C$_{2}$F$_{4}$}
\label{figure_structures}
\end{figure}

\begin{figure}
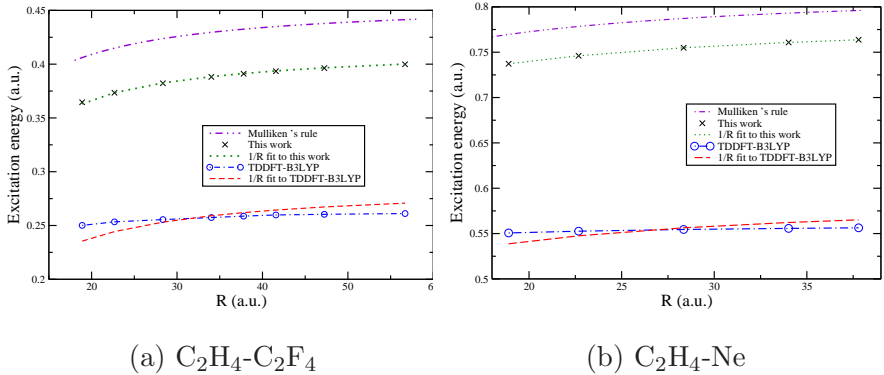

\centering
\begin{subfigure}[b]{0.35\textwidth}
\includegraphics[width=\textwidth]{C2H4C2F4_R.eps}
\caption{C$_{2}$H$_{4}$-C$_{2}$F$_{4}$}
\end{subfigure}
\begin{subfigure}[b]{0.35\textwidth}
\includegraphics[width=\textwidth]{C2H4Ne_R.eps}
\caption{C$_{2}$H$_{4}$-Ne}
\end{subfigure}
\caption{Charge-transfer energy at different intermolecular
separations $R$. Apart from the energies calculated from the approach
presented here, we give for comparison the fit of these values to a 1/R
curve, the curve that one gets from Mulliken's rule as well as the
TDDFT-B3LYP curve with its own 1/R fitted curve.}\label{figure1}
\end{figure}

The charge-transfer excitations at large intermolecular distances follow the
Mulliken's rule\cite{Mulliken} which states that the lower photon energy $%
E_{CT}$ required to induce an electron transfer between a donor - acceptor
system in asymptotically large intermolecular distances $R$ is: 
\begin{equation*}
E_{CT}=IP^{D}-EA^{A}-1/R,
\end{equation*}%
where $IP^{D}$ is the ionization potential of the donor and $EA^{A}$ is the
electron affinity of the acceptor. To check whether the energy grows
asymptotically with 1/R, we studied the lowest charge-transfer energy of the two
systems C$_{2}$H$_{4}$-C$_{2}$F$_{4}$ and C$_{2}$H$_{4}$-Ne at different large intermolecular distances $R$ (see Fig.\ref{figure_structures}) and
we present the results in Fig.\ref{figure1}. For comparison, we plot the curve we get from Mulliken's law, the fitted curve to our
points where energy grows with 1/R and the curve derived from
TDDFT-B3LYP with its 1/R fit. For both complexes the aug-cc-pVDZ basis set
was used for all the energy calculations. Although in both cases, our
results are below the ones from Mulliken's rule, they satisfy the
1/R behavior contrary to the TDDFT-B3LYP results that apart from being too low
in energy are almost constant with $R$.\par

For all the calculations with our minimization scheme we used the ground state HF orbitals that
correspond to the $S^{A}_{occ}$ subspace instead of solving
the eigenvalue equation (\ref{eq:eigenstate acceptor2}).
This is due to the fact that optimizing this set of orbitals gave only a minor
change to the total energy, much smaller than the accuracy of our results.
Nevertheless, we cannot claim that one can always omit solving this
equation, although it is expected that in this subspace the
changes with respect to the ground state orbitals are smaller than in the other two, $%
S_{vir}^{A}$ and $S_{occ}^{D}$, where the particle and the hole are located.\par

For all of our calculations we used basis sets with diffuse
functions. Those functions are particularly important when  the donor and acceptor subsystems
are not close to each other to ensure that one treats two weakly
interacting systems and not two isolated ones. All the ground state and the TDDFT-B3LYP calculations for
the C$_{2}$H$_{4}$-Ne C$_{2}$H$_{4}$-C$_{2}$F$_{4}$ systems were carried out using Gamess US \cite{gamess}
and a code was developed for our approach. For the visualization of the structures 
the xcrysden program was used \cite{xcrysden}.

\section{CONCLUDING REMARKS}

In this work we extend our HF excited state subsequent
minimization scheme developed recently to describe charge-transfer
excitations. This is achieved by creating a particle-hole pair excited
state, which is always orthogonal to the ground state and both the particle
and the hole are determined variationally. This is done by separating the
ground state HF orbitals into two sets: one mostly localized at the donor
and the other at the acceptor. Each such set is further separated in 
subspaces of occupied and virtual orbitals. Thus, we have four subspaces and
the HF energy functional is minimized in each subspace separately. In this
way variational collapse and self-interaction are avoided.\par
 We tested the approach on a standard set of pairs of molecules and found that our
theoretical results are in good agreement with the experimental ones and
show the correct asymptotic behavior. It is worth mentioning that the
computational cost of this scheme is the same as the one of a HF ground
state calculation. Finally, one could easily calculate double or multiple
excitations within this approach as the only difference is that
one would need to create more than one particle-hole pair in the
appropriate subspaces. 

\section{AKNOWLEDGMENTS}
The authors would like to thank Dr. N. Helbig, Dr. N.N. Lathiotakis and Prof. A.K. Theophilou 
for usefull discussions. I.T. acknowledges financial support from the DFG through the Emmy-Noether program.

\end{document}